\documentclass[journal=jacsat,manuscript=article]{achemso}
\usepackage[version=3]{mhchem}

\author{Sergii Morozov}
\affiliation[Imperial]
{The~Blackett~Laboratory, Department~of~Physics, Imperial~College~London,~London~SW7~2AZ, United~Kingdom}
\email{s.morozov@imperial.ac.uk}

\author{Michele Gaio}
\affiliation[Imperial]
{The~Blackett~Laboratory, Department~of~Physics, Imperial~College~London,~London~SW7~2AZ, United~Kingdom}

\author{Stefan A. Maier}
\affiliation[Imperial]
{The~Blackett~Laboratory, Department~of~Physics, Imperial~College~London,~London~SW7~2AZ, United~Kingdom}
\alsoaffiliation[LMU]
{Chair in Hybrid Nanosystems, Faculty of Physics, Ludwig-Maximilians-Universit\"at~M\"unchen, 80799~M\"unchen, Germany}

\author{Riccardo Sapienza}
\affiliation[Imperial]
{The~Blackett~Laboratory, Department~of~Physics, Imperial~College~London,~London~SW7~2AZ, United~Kingdom}
\email{r.sapienza@imperial.ac.uk}
\phone{+44 (0) 207 594 9577}

\title[A metal-dielectric parabolic antenna to direct single photons ]
  {A metal-dielectric parabolic antenna to direct single photons }

\abbreviations{QD,AFM,FDTD,NA,APD}

\usepackage[normalem]{ulem}

\usepackage{multicol}
\usepackage{titlesec}
\titleformat*{\section}{\small\bfseries}

\makeatletter
\let\l@addto@macro\relax 
\makeatother
\usepackage[fontsize=11pt]{scrextend}
\usepackage{multirow}

\begin{document}
\setstretch{1.0}
% \begin{tocentry}

% \centering

% \includegraphics{fig0}
% \label{For_TOC_only}

% \end{tocentry}

\begin{abstract}
Quantum emitters radiate light omni-directionally, making it hard to collect and use the generated photons. 
Here we propose a 3D metal-dielectric parabolic  antenna surrounding an individual quantum dot as a source of collimated single photons which can then be easily extracted and manipulated.
Our fabrication method relies on a single optically-induced polymerization step, once the selected emitter has been localized by confocal microscopy.
Compared to conventional nano-antennas, our  geometry does not require near-field coupling and it is therefore very robust against misalignment issues, and minimally affected by  absorption in the metal. 
The parabolic antenna provides one of the largest reported experimental directivities ($D=106$) and the lowest beam divergences ($\Theta_{1/2}=13.5^\circ$), a broadband operation over all the visible and near-IR range, together with more than 96\% extraction efficiency, offering a practical advantage for quantum technological applications. \\

\centering
\includegraphics{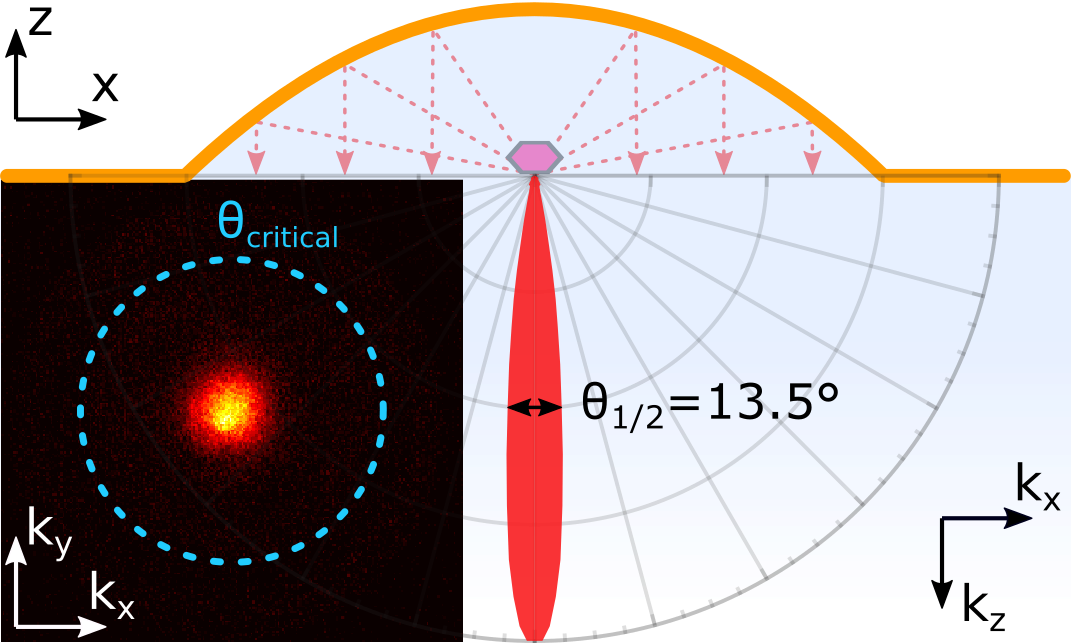}
\end{abstract}

%{{Keywords: optical antenna, nanophotonics, quantum dot, single photon emitter, directional emission, directivity, direct laser writing.}}\\

\clearpage
\begin{multicols}{2}

A single photon can be used for storage and long-distance transport of quantum information.
Single photon sources have been realized in various solid state systems, such as single molecules \cite{Basch1992}, color centers in diamond \cite{Kurtsiefer2000}, defects and color centers in 2D materials \cite{He2015,Tran2015}, quantum dots \cite{Michler2000}. However, practical implementations of these sources for single-photon technologies are limited by their omni-directional radiation, which makes it hard to collect and use the generated photons. 

Plasmonic nano-antennas \cite{Novotny2011,agio2013optical} have shown an elegant solution to this problem, as photons can be extracted by near-field coupling to metallic resonant elements, and radiated to the far field. 
Although a variety of  plasmonic antenna design have been proposed, ranging from nanoaperture \cite{Aouani_2011}, optical patch antenna \cite{Belacel_2013}, v-shaped antennas \cite{Vercruysse2014} and Yagi-Uda antenna \cite{Curto2010}, they are all constrained by the large Ohmic losses in the metal which limit the single-photon extraction efficiency.
Recently, dielectric antennas sustaining Mie resonances have attracted much interest as they feature low losses and a potential for quantum emitters engineering \cite{Kuznetsov2016,Cambiasso2017}. 
Nevertheless, both plasmonic and dielectric nano-antennas feature a limited directivity, which is constrained by their subwavelength size \cite{Koenderink2017}.

If the requirement of near-field coupling and sub-wavelength sizes is relaxed, a much larger directivity and extraction efficiency can be achieved, as shown for example for a planar dielectric-metal antenna \cite{Chu_2014}, circular bullseye shaped antennas \cite{Livneh2016}, or a silver coated nanopyramidal structure \cite{Kim_2016}. Parabolic shaped nano and micro-structures have been demonstrated to act as unidirectional and broadband antennas at optical frequencies \cite{Schoen2012,Schell2014}.

For all the designs mentioned above, a common practical challenge lies in the precise emitter-antenna interfacing required for their operation. Several techniques for the deterministic positioning of an emitter near an antenna have been developed, such as  multi-step lithography in combination with chemical functionalization \cite{Curto2010}, nano-manipulation by the tip of atomic force microscope \cite{Schietinger_2009,Andersen2018}, near-field polymerization \cite{Volpe2012}, and optical trapping \cite{Geiselmann2014}. However, a controlled emitter-antenna interfacing remains rather a complex in-lab technology with poor reproducibility, and therefore an opposite approach, to bring an antenna over an emitter, has been pursued for example using \emph{in-situ} far-field optical lithography \cite{Dousse2008}{, post-alignment electron-beam processing \cite{LSapienza2015,Schlehahn2016}, and a combination of \emph{in-situ}
electron-beam lithography and optical 3D laser writing \cite{Fischbach2017}}. 

%In this Letter
In this Letter, we report a 3D hybrid metal-dielectric parabolic antenna to control the  emission direction of an individual colloidal quantum dot. 
The design and fabrication circumvents the emitter-antenna alignment issues,  because the antenna is made \emph{in-situ} over a selected individual quantum dot.  
By optical Fourier imaging we observe a redirection of the emitted single photons into a beam in the direction normal to the sample plane with the divergence of $13.5^\circ$, and over 20-fold increase of directivity comparatively to a quantum dot at air-glass interface. As a result, the  single photon extraction rate is predicted to exceed 96\% for $\text{NA}=1.5$ and 60\% for $\text{NA}=0.5$, while allowing for a broadband operation over the full visible and near-IR range.  

\begin{figure*}
  \includegraphics[width=0.98\textwidth]{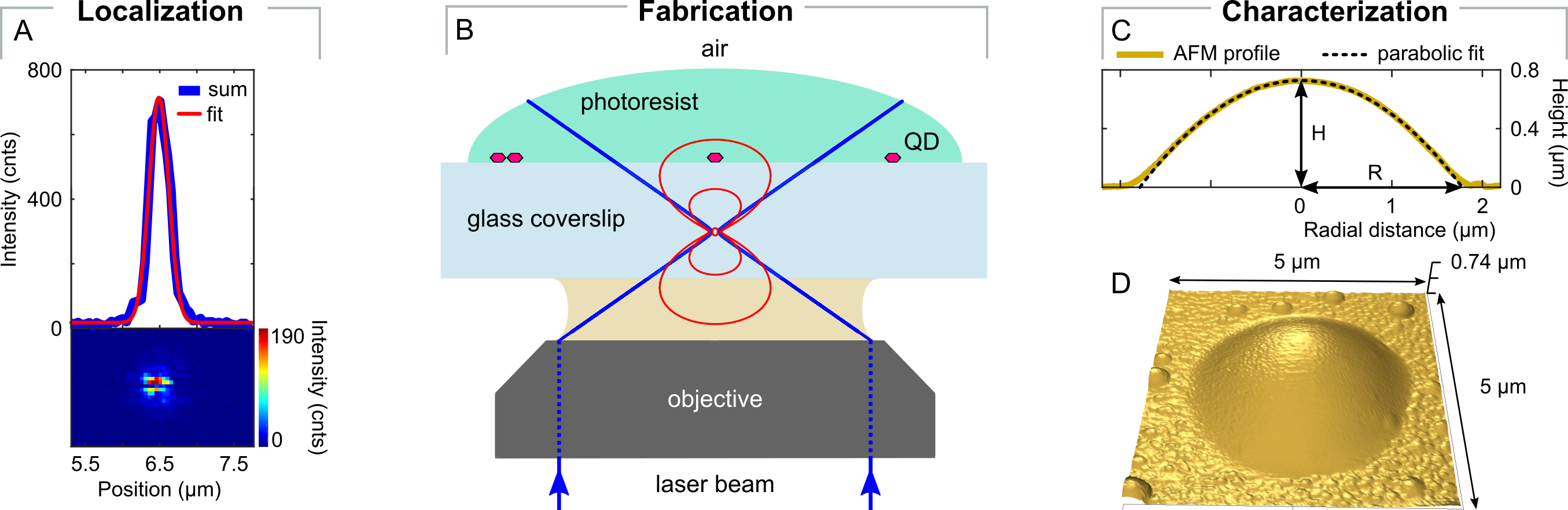}
  \caption{ \small \textbf{Fabrication and characterization of a parabolic antenna interfaced with a single quantum dot (QD).} (A) Localization of a single quantum dot by fluorescence confocal microscopy with {sub 10~nm precision}. Bottom panel presents a confocal scan of a single quantum dot on a glass coverslip, and the top panel demonstrates an integrated profile of the confocal image (blue) and its Gaussian fit (red). (B) Schematic of the parabolic antenna fabrication (not to scale): a blue laser beam (blue) is focused below the glass-photoresist interface in order to polymerize the volume of photoresist confined by the weakest intensity isoline (red) around the target quantum dot. (C)~The AFM profile of a parabolic antenna (dark yellow) is fitted with a parabolic function (dashed black). (D) An AFM scan of a fabricated parabolic antenna on a~substrate.}
\label{fgr:one}
\end{figure*}

%[Fabrication]\\
The proposed parabolic antenna is inspired by macroscopic parabolic reflectors, which project electromagnetic radiation from the focal point into a beam parallel to the parabola axis. 
We scale this concept to the micrometer size, and explore how far it can be applied to collimate the single-photon emission of a  quantum emitter. 
The parabolic antenna is fabricated top-down on an individual emitter in a single-step of laser-induced polymerization. The fabrication scheme is the following:
(1)  a colloidal CdSeTe/ZnS quantum dot on a glass coverslip is localised by confocal microscopy, as in Fig 1A, with a typical sub 10~nm precision {(which is the localization error of the centre of a diffraction limited spot \cite{Balzarotti2016})}, (2) a polymeric paraboloid is fabricated around it  by one-photon polymerization of IPL-780 photoresist. 
This method exploits the parabolic shape of the top-most part of the iso-intensity profile of a focused Gaussian laser beam (see Fig.~\ref{fgr:one}B, red closed lines)  to expose the paraboloid in a single 70~ms laser illumination step at the exposure power of 2~$\mu$W. 
By controlled defocussing of the writing beam,  (see Fig.~\ref{fgr:one}B), the paraboloid is aligned with the target quantum dot in its focal point.
After washing the unexposed part of the photoresist, the resulting parabolic antenna is revealed. Atomic force microscopy (AFM) scan in Fig.~\ref{fgr:one}D and its cross-section in Fig.~\ref{fgr:one}C  confirm that the polymerized paraboloid follows  the target shape with a surface deviation from the parabolic shape of $\pm 40$~nm (see Supporting Information). The alignment of the antenna with the target quantum dot is confirmed post-fabrication by confocal imaging (see Supporting Information).
Finally, (3) the polymeric antenna is covered with a thin 30-50~nm layer of gold to increase its reflectivity. 
The presence of the metal does not come with a substantal increase of photon losses, as the target quantum dot is about a wavelength (800~nm) away from the gold layer, too far for near-field coupling to the metallic shell.

The structural dimensions of the parabolic antenna can be controlled during the fabrication by two parameters, the exposure dose and the defocussing distance, which determine the size and curvature of the antenna.
%. 
In order to achieve the collimated emission, the target quantum dot resting on a glass coverslip has to be in the focal point of the antenna, which is achieved when the paraboloid height $H$ and the radius $R$ of its aperture are related as $R=2H$~(Fig.~\ref{fgr:one}C). 
In that configuration  the paraboloid focal length is  $f = H$. 
As a compromise between size and directivity, we have chosen the focal length  $f$ to match the quantum dot emission maximum wavelength $\lambda_{\text{em}}=800$~nm, thus the fabricated antenna has the geometrical parameters $H=\lambda_{\text{em}}$ and $R=2\lambda_{\text{em}}$, as~shown~in~Fig.~\ref{fgr:one}C {(see also Supplementary Information)}.

\begin{figure*}
  \includegraphics[width=0.49\textwidth]{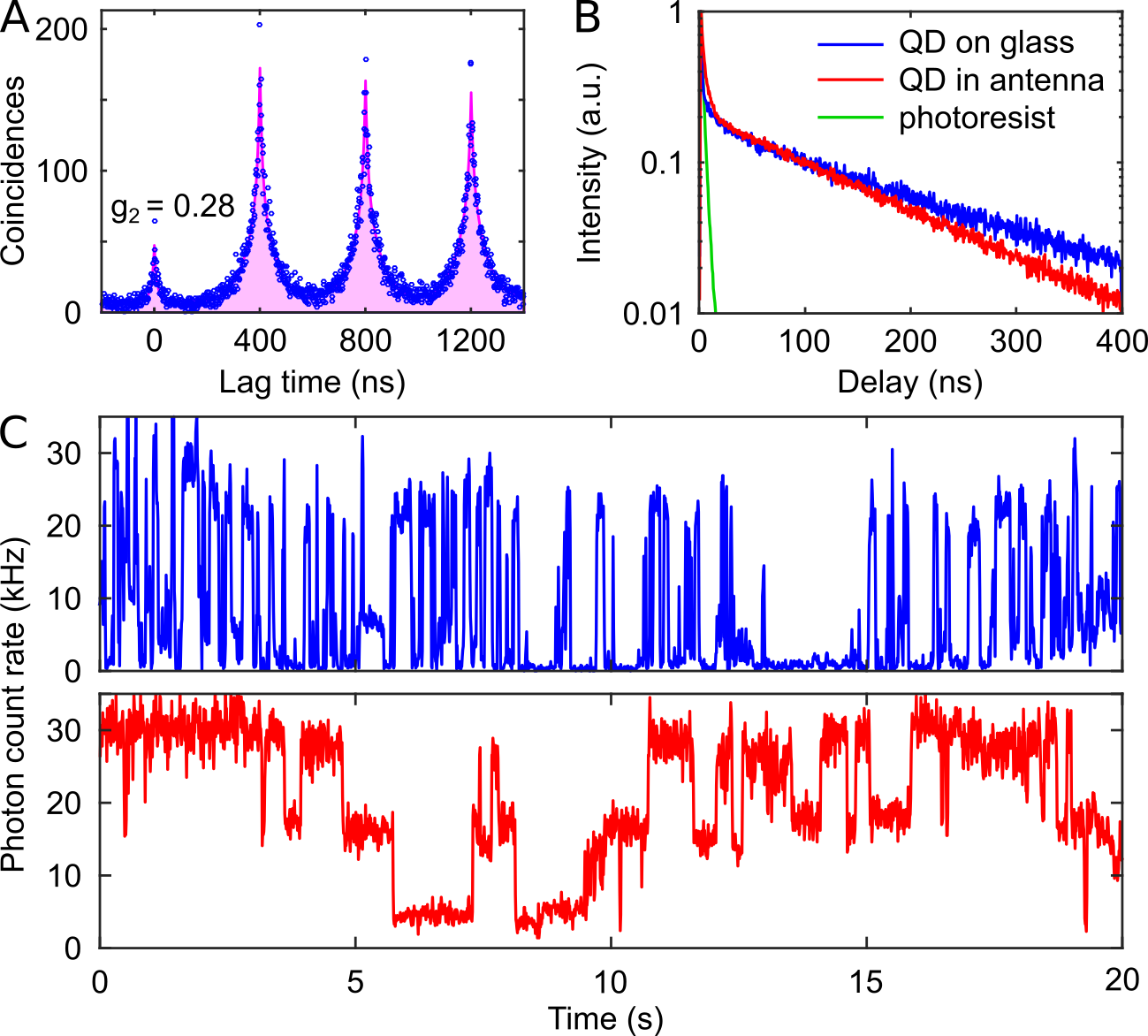}
  \caption{\small \textbf{Time dynamics of a single quantum dot emission.} (A) Measured second-order intensity correlation function (blue), and its fit (magenta). (B) Normalized decay histograms of the target quantum dot (QD) are shown before (blue) and after (red) embedding QD into the antenna. The decay histogram of photoresist (green) was measured from an antenna without a QD. (C) Intensity time traces (time bin is 10~ms) of the target QD are measured with excitation power of 18~nW at 2.5~MHz before (blue) and after (red) embedding QD into the antenna.}
\label{fgr:two}
\end{figure*}

The parabolic antenna is  expected to substantially leave the local density of optical states unaltered as it does not sustain optical resonances, with only a minor increase of Purcell factor of less than 30\% revealed by {numerical simulations} (see Supplementary Information), related to the change of the quantum dot environment from an air-glass to an almost index-matched polymer-glass interface.
Fluorescence dynamics studies in Fig.~\ref{fgr:two}B  shows that while the overall decay histogram of the target quantum dot before and after the fabrication is multi-exponential, the long decay component, attributed to neutral excitons, shortens from 159~ns to 132~ns when the emitter is covered with the photoresist. 
The short lifetime \textless~5~ns components arise from multi-exciton excitation and Auger recombination processes  \cite{Sampat2015,Galland_2012,Ox-Malko2011}. 
As shown in Fig.~\ref{fgr:two}A, before measuring the emission dynamics we test the fluorescence signal with an Hanbury-Brown and Twiss interferometer, to confirm antibunching of the emitted photons. For the specific case shown in  Fig.~\ref{fgr:two}A we report the second-order coherence  $g_2=0.28$.
The photoresist is weakly luminescent with emission lifetime around 2.5~ns, which we report for completeness as the green decay histogram in Fig.~\ref{fgr:two}B, and which contributes to the fast decay at short times. 
The complex blinking behavior shown in Fig.~\ref{fgr:two}C prevents a precise assessment of the emission saturation values, but indicates that the maximum count rate is around 30~kCounts/s, which is preserved after the antenna fabrication. 
Fig.~\ref{fgr:two}C indicates also that the off-times in the intensity time trace are slightly less probable in a quantum dot interfaced with the antenna, which we attribute to the protection of the target quantum dot from oxygen atmosphere when capped by the polymer, and the change in electrostatic environment \cite{Tang_2005,Ko_2010,Ox-Qin2012}. 
As a result, the total angular-integrated emission count rate averaged over a minute grows from 8.5~to~23.9~kCnts/s when the quantum dot is interfaced with the antenna. 

\begin{figure*}
  \includegraphics[width=0.98\textwidth]{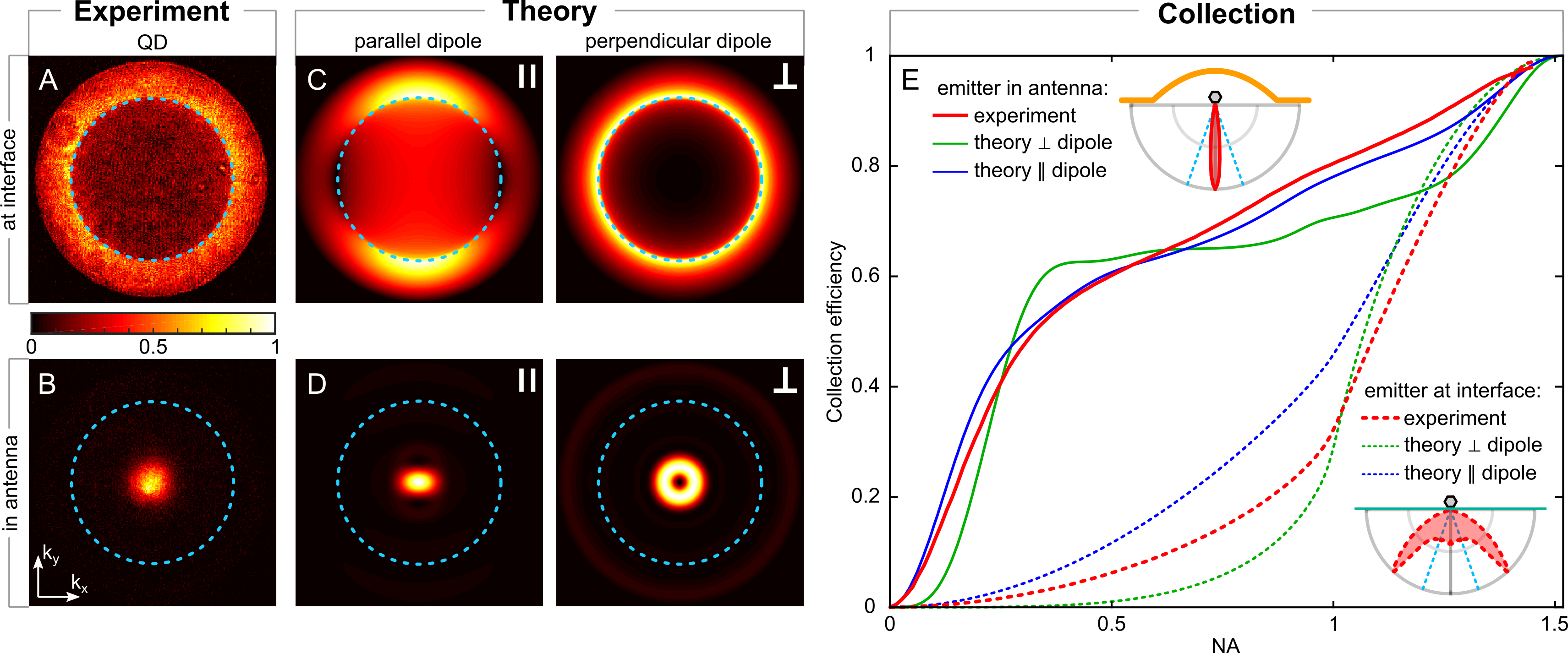}
  \caption{ \small \textbf{Experimental and theoretical far-field radiation patterns.} Measured (A,B) and simulated (C,D) angular radiation patterns of an emitter at air-glass interface (top row), and in the focal point of the parabolic antenna (bottom row). 
(B) A single quantum dot in a parabolic antenna has a radially symmetric radiation pattern with the main lobe centered at $\Theta=0^\circ$ and the half power beamwidth of $\Theta_{1/2}^{\text{exp}}=13.5^\circ$. The simulated radiation patterns in (C) and (D) are shown with respect to the parallel and perpendicular dipole orientations. The dashed blue circles show the air-glass critical angle.  
(E) Collection efficiency as a function of the numerical aperture ($\text{NA}$) of an objective. The set of dashed lines shows the dependence for the case of an emitter at air glass interface, while the set of solid lines demonstrates the improvement of collection efficiency for an emitter interfaced with the parabolic antenna. The insets in (E) present radial profiles of angular radiation patterns from (A)-(B) in polar coordinate system (red). The dashed magenta lines show the $\text{NA}=0.5$ opening. }
\label{fgr:three}
\end{figure*}
%[Results: Angular emission]
The antenna strongly affects the angular radiation of an emitter placed in the focal point of the parabolic antenna. 
This is evident when we compare the angular radiation patterns of an emitter before and after embedding it into the antenna. 
The experimental radiation pattern of a quantum dot at the air-glass interface (Fig.~\ref{fgr:three}A)  features a rotational symmetry pattern with most of the light outside the air-glass critical angle (blue dashed circle).
This is consistent with the 2D degenerated transition dipole moment of a quantum dot, resulting into an emission pattern which is the sum of individual dipolar patterns \cite{Koberling}.
Instead, when the quantum dot is interfaced with the parabolic antenna, as presented in Fig.~\ref{fgr:three}B, a sharp redirection of the emission occurs into a narrow beam normal to the sample plane with the half power beamwidth of $\Theta_{1/2}^{\text{exp}}=13.5^\circ$. 
This beam is inside the light cone and can be collected without high NA optics due to its low divergence.   
These experimental results are well confirmed by numerical simulations (Fig.~\ref{fgr:three}C-D).
In the simulations shown in Fig.~\ref{fgr:three}C-D, we can distinguish the contributions from the two dipole polarizations, parallel~($\parallel$) and perpendicular~($\perp$) to the sample plane. 
Therefore, we can highlight that the parallel dipole is collimated around the normal direction with the emission divergence of $\Theta_{1/2}^{\parallel}=10.9^\circ$, while the perpendicular dipole has a donut-like shape with a minimum in the normal direction and an annular emission very close to it. 
This is what one would expect from simple geometrical reasoning for a dipolar radiation. 

The redirection of emission shown in Fig.~\ref{fgr:three}B and D results in improved collection efficiency. 
{We calculated the collection efficiency as a function of $\text{NA}$ (Fig.~\ref{fgr:three}E) by measuring the emission pattern with $\text{NA}=1.45$ optics and selecting a section of it by radial integration of the experimental and theoretical angular radiation patterns (Fig.~\ref{fgr:three}A-D).
This is the relative collection efficiency which neither includes the imperfect detection efficiency of our setup, which we estimate to be around 5-10\%, nor the minor angular-dependent detection efficiency.}
That is, photons originally emitted towards bottom half space are not affected by the parabolic antenna, and, therefore, cannot be collected with a low $\text{NA}$ objective. 
For a quantitative comparison, we analyze emission from the emitter in $\text{NA}=0.5$ opening towards a detector. 
We achieved experimentally a 10-fold increase of the collection efficiency with an $\text{NA}=0.5$ objective of single photon emission from a quantum dot, namely from $6\%$ to $61\%$.

The angular collimation reported in Fig.~\ref{fgr:three}B and D in the $k_x$-$k_y$ momentum space can be also described using the antenna directivity $D$ \cite{NovotnyBook,Balanis2015}. 
It is defined as a ratio of the radiation intensity in a given direction to the averaged radiation intensity in all directions. For the current design we achieved the directivity of $D_\text{exp} = 106$.
This value of directivity compares well to the  prediction from  numerical calculations  for a dipole in the focal point of the parabolic antenna (Fig.~\ref{fgr:three}D), which give $D_{\parallel}= 151$ and $D_{\perp} =70$ for the parallel and perpendicular dipole, respectively.
This is to our knowledge one of the largest reported directivities, surpassing that of metal and dielectric antennas of similar size, which is in the range 5-25 \cite{Lee_2011,Aouani2_2011,Schoen2012,Peter2017}.

\begin{figure*}
\includegraphics[width=0.49\textwidth]{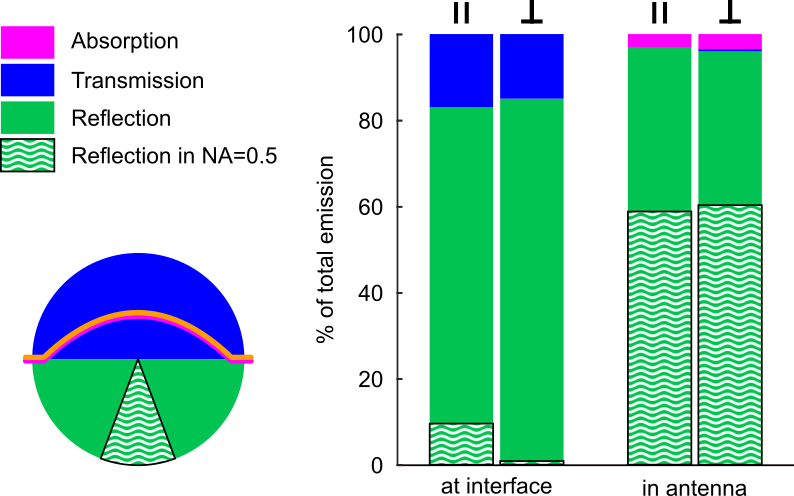}
\caption{ \small \textbf{The distribution of total  power emitted by a dipole.} The numerical simulations of the parallel~($\parallel$) and perpendicular~($\perp$) dipoles demonstrate the distribution of the total emitted power before and after the emitter-antenna interfacing, which is allocated to absorption, transmission and reflection. The diagram on the left shows schematically the upper (blue, transmission) and lower (green, reflection) half-spaces, as well as the $\text{NA} = 0.5$ opening towards the detector (wavy-white and green segment).}
\label{fgr:four}
\end{figure*}

The parabolic antenna reflects the light which otherwise would have been lost into the upper half-space, and therefore increases the collection from the glass side.
This can be quantified theoretically by calculating the photon extraction rate, and by analyzing how much of the emitted light is absorbed, transmitted and reflected by the quantum dot-antenna system.
{The numerical simulations plotted in Fig.4 show that in absence of the antenna, about $\eta^{1.5}_{\parallel} =83$\% and $\eta^{1.5}_{\perp}=85$\% (depending on the dipole orientation) of the total emitted power goes to the lower half-space, i.e. in NA = 1.5, whereas the rest part of emission is lost in upper halfspace. 
Instead, if the emitter is interfaced with the parabolic antenna, $\eta^{1.5}_{\parallel} = 97$\% and $\eta^{1.5}_{\perp} = 96$\%  of the emitted photons can be extracted in the lower half-space, with a few percent absorption in the gold shell.}
This small increase in  extraction efficiency for $\text{NA} = 1.5$ becomes a very large for $\text{NA} = 0.5$, as this efficiency is drastically boosted by the directional radiation (wavy-white and green bars in Fig.~\ref{fgr:four}). 
At the air-glass interface the power extraction rates  in $\text{NA} = 0.5$ solid angle from a bare dipole are limited to $\eta^{0.5}_{\parallel} = 12\%$ and $\eta^{0.5}_{\perp} =1\%$, while the antenna improves these values to $\eta^{0.5}_{\parallel} = 61\%$ and $\eta^{0.5}_{\perp} = 63\%$, providing over a 5- and 60-fold increase of the extraction rate, respectively. 
The simulations show that absorption in the gold layer accounts for less then $4\%$ of the total power loss (magenta bars in Fig.~\ref{fgr:four}), while transmission through the parabolic reflector is completely suppressed~(\textless$1\%$). 
 
\begin{figure*}
  \includegraphics[width=0.98\textwidth]{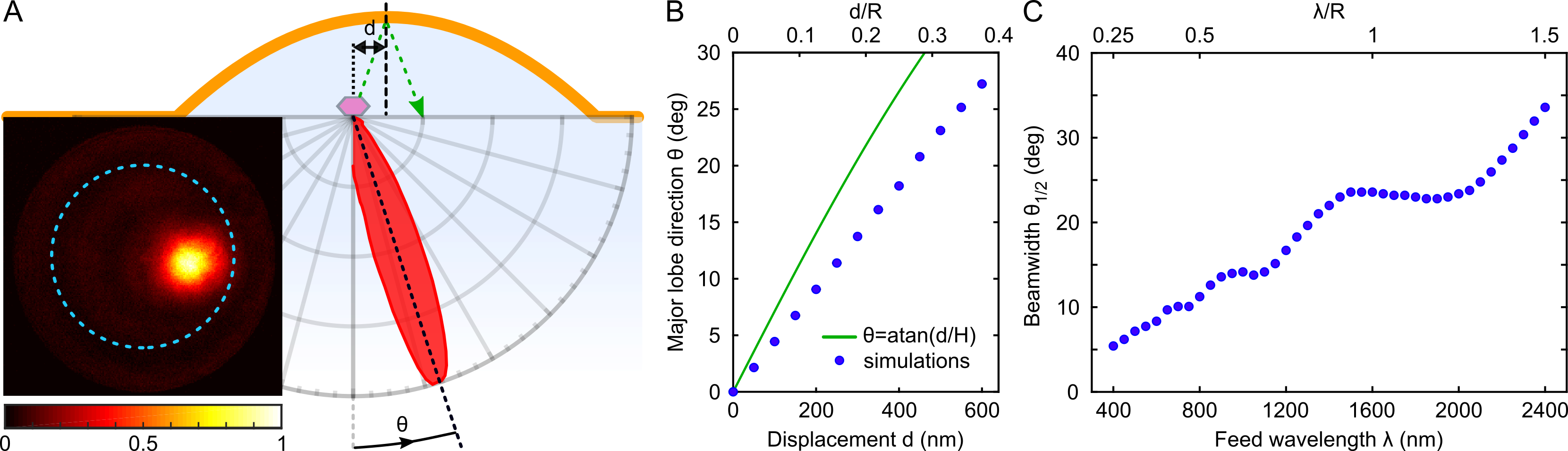}
  \caption{ \small \textbf{Operational limits of the parabolic antenna.} (A) The experimental angular radiation pattern of a quantum dot displaced from the parabola axis by $d$ demonstrates the tilt of the collimated beam $\Theta$, and the experiment schematics with a radial profile in polar coordinate system of the measured angular radiation pattern. The blue dashed circle shows the air-glass critical angle. (B)~The dependence of the collimation direction $\Theta$ on the lateral displacement $d$ of the feed. The green curve shows the dependence obtained from a simple reflection calculation, and the blue dots show results of numerical simulations. (C) Simulated dependence of the beam divergence $\Theta_{1/2}^\parallel$ of a parallel dipole placed in the focal point on the wavelength $\lambda$ for the antenna with $R=1600$~nm aperture.}
\label{fgr:five}
\end{figure*}

The antenna provides unidirectional single photon emission if the emitter is placed in its focal point. 
Radial and axial displacements of the emitter from the ideal position affect its radiation pattern by  tilting the emerging beam and by widening the emission angle. 
We have observed that quantum dots placed below and above the focal plane have angular emission patterns with wider beamwidth and parasitic minor lobes, thus with a lower directivity (see Supporting Information).  
When an emitter is on the focal plane but displaced sideways by a distance $d$ from the axis of symmetry of the antenna, its emission remains directional, but it is tilted  with respect to the normal to the sample plane as shown experimentally in Fig.~\ref{fgr:five}A. 
The tilt can be explained in the approximation of geometrical and ray-optics considerations, that is the green ray shown in Fig.~\ref{fgr:five}A  which gives us a simple relation between the displacement $d$ and the direction of the major lobe $\Theta=  \text{atan} (d/H)$ (green curve in Fig.~\ref{fgr:five}B).
However, this approximation does not take into account the radiation pattern of the feed and the coma aberration introduced by the paraboloid shape of the antenna \cite{BornWolf}, so the actual tilt is smaller as confirmed by numerical simulations of angular radiation pattern of a parallel dipole displaced from the focal point (blue points in Fig.~\ref{fgr:five}B).
The experimental angular radiation pattern shown in  Fig.~\ref{fgr:five}A has the tilt of  major lobe of $\Theta=19^\circ$ to the normal, which is predicted to correspond to the displacement of about $d= 400$~nm from the aperture center.
Therefore,  our geometry is very robust to radial misalignment of the target quantum dot with antenna, and it  can be also used to direct single photons at specific angles up to $30^\circ$.
Given our sub 10~nm experimental precision in placing the antenna on the target quantum dot, the error in the direction of the major emission lobe is kept below~$1^\circ$. 

The parabolic antenna does not sustain any optical resonances, therefore its response is broadband, with the only wavelength dependence in the parabola collimation power. The ability of the antenna to collimate light follows a simple Fourier relation with the antenna size: the larger antenna the lower the beam divergence.  That is the beam divergence scales with the $\lambda/R$ ratio \cite{Balanis2015}. This prediction can be tested  by numerical simulations of the beam divergence $\Theta_{1/2}$ evolution versus the feed wavelength, as shown in Fig.~\ref{fgr:five}C for the case of a parallel dipole. In Fig.~\ref{fgr:five}C a linear dependence of $\Theta_{1/2}$ on $\lambda/R$ is visible, as expected, with  an over-imposed oscillation which we attribute to the opening of parasitic side-lobes, which are especially strong at large $\lambda/R$ ratios. Instead, for large parabolic antennas, i.e. small $\lambda/R$ ratios, the oscillations are reduced and the general behavior follows the linear scaling of $\Theta_{1/2} \propto \lambda/R$ also found in macroscopic parabolic antennas. The dotted curve  in Fig.~\ref{fgr:five}C shows that the same parabolic antenna design can collimate radiation spanning the visible as well as telecom spectral range, with only a  drop in the beam divergence. 

{The fabrication process is very versatile and can be adapted to various quantum emitters and antenna shapes, albeit limited to semi-planar geometries, sizes bigger than diffraction limit and below 10 micrometers due to laser power requirements. Most resists are also fluorescent which could impeded single-photon operation.}

%[Conclusions]
In conclusion, we have presented a 3D parabolic antenna fabricated with a simple photo-polymerization step capable of directing single photon emission into a narrow beam with divergence of $\Theta_{1/2}^{\text{exp}}=13.5^{\circ}$. This corresponds to one of the highest directivities $D=106$ of single photon emission reported so far, and results in a more than 96\% total and 60\% in $\text{NA} = 0.5$ extraction efficiency.  This can be very helpful for coupling into a fibre or for photon extraction in cryogenics conditions. The laser writing processing does not damage the emitters, and it can be flexibly applied to a wide range of single photon emitters, e.g. from defects in diamonds to individual molecules. The emitter-antenna hybrid delivers high-directional photons, with \textless$4\%$ absorption losses, and it is compatible with integration with nano-antennas and near-field photon engineering, offering a practical advantage for optical quantum technologies.

\textbf{Methods.} The experimental setup is a custom-built time-resolved confocal fluorescence microscope. It is based on the body of Nikon eclipse Ti inverted microscope. Samples were scanned with a 3D piezo stage (E-545.3CD PI nano). A large NA oil-immersion objective (Plan Apochromat 100x, NA=1.45) focused the laser radiation on a sample and collected the photoluminescence of a quantum dot. The collected photoluminescence signal was separated from the excitation laser by a dichroic mirror, and further filtered with a $795\pm90$~nm bandpass filter, and either sent to a grating spectrometer (IsoPlane SCT 320) with a CCD camera (PIXIS 400B eXcelon) or to avalanche photo diodes (APDs, SPCM-AQRH, Perkin Elmer).
 
A blue laser (LDH-D-C-440 Picoquant) at 442~nm with pulse width 64~ps and repetition frequency 2.5~MHz was used to excite quantum dots. Antibunching curves were measured using Hanbury-Brown and Twiss setup. It consists of  two APDs which are connected to a time-correlated single photon counting module (TimeHarp~260, PicoQuant). The arrival times of photons on the detectors are collected into a histogram, producing the second order correlation function to verify the single photon emission. Decay histograms were collected by recording the time between the laser excitation pulse and the arrival time of a photon on a single APD.
Angular radiation patterns of quantum dots were measured by imaging the back-focal (Fourier) plane on the CCD camera. 

The theoretical model is based on the numerical solution of the Maxwell’s equations with a finite difference time domain (FDTD) method. The 3D FDTD simulations were performed using a commercial package (Lumerical).

\begin{acknowledgement}
The authors wish to thank Avi Braun and Sandro Mignuzzi for fruitful discussions. This research was funded by the Engineering and Physical Sciences Research Council (EPSRC EP/P033431, EP/M013812). S.M. acknowledges the Cromwell scholarship and S.A.M. the Lee-Lucas Chair in Physics.

\end{acknowledgement}

\begin{suppinfo}
Details on the design, fabrication and characterization of parabolic antennas can be found in the Supporting Information.
\end{suppinfo}
 
 \end{multicols}
\bibliography{Bibliography}

\end{document}

\setstretch{1.2}
 %\section{Introduction}
 \paragraph{Sample fabrication.} 

%\textbf{Sample fabrication.} 
We used commercially available Qdot 800 ITK\texttrademark  \space organic quantum dots. The quantum dots were deposited on a microscope glass coverslip by spin coating. Parabolic antennas were fabricated by one photon polymerization of drop casted photoresist on glass coverslips. 
We used a commercially available photoresist IPL-780. The exposed samples were placed first in a PGMEA bath for 15 minutes, and then in an isopropanol bath for 5 minutes to remove the unexposed photoresist. Afterwards, the samples were covered with a 30-50 nm gold layer using Q150T turbo-pumped sputter coater. We used Bruker Icon Dimension atomic force microscope for the geometrical characterization of the fabricated antennas. 
\paragraph{Parabolic antenna design.}
The shape of a fabricated structure is defined by the weakest isosurface of intensity able to polymerize photoresist. The samples have been aligned, placing  the writing focused beam over each  quantum dot  by a piezo stage. We polymerized a parabolic structure using the laser intensity profile whose iso-intensity lines are solutions to the equation:
\begin{equation}
   I(r,z) = I_0 \Big(\frac{w_0}{w(z)}\Big)^2 e^{-\frac{2r^2}{w(z)^2}},
\label{eqn:isolines}
\end{equation}
where $r$ and $z$ are radial and axial distances respectively, $w_0$ is the waist radius defined by the Abbe diffraction limit, $w$ is the spot size parameter, and $I_0$ is the intensity at the center of the beam. As an example, the solutions for 3 arbitrary intensities are shown in Fig.~1B as closed red curves; the strongest intensity is enclosed in the smallest volume. 

\paragraph{Characterization of fabricated parabolic antennas.}
 
An optical microscope image of 6 fabricated parabolic antennas around preselected single quantum dots is shown in Fig.S1A. 
A confocal scan of an area shown by a green square in Fig.S1A, verifying antenna-emitter alignment, is demonstrated in Fig.S1B. 
The intensity blinking of a single quantum dot can be seen in the confocal scan as dark lines in the middle of the image. 
A 3D parabolic fit of the antenna from Fig.S1B shows the deviation from ideal paraboloid shape, which is in range $\pm 40$~nm.

\begin{figure}
\includegraphics[width=0.96\textwidth]{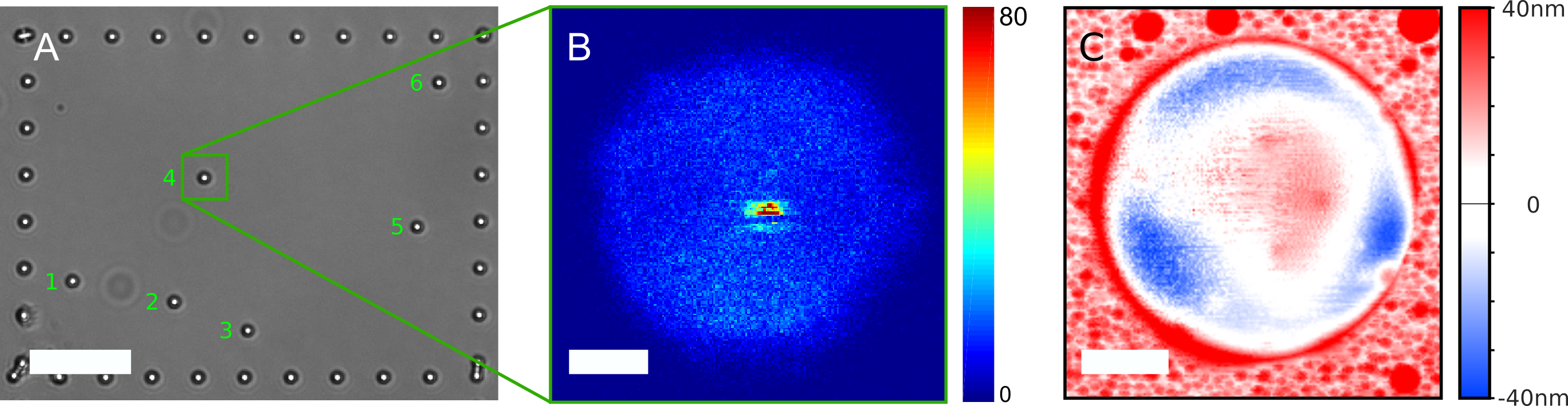}
\caption{\textbf{Fabrication and characterization of parabolic antennas.} (A)  Parabolic antennas 1-6 (green) edged by markers. The scale bar is 10~$\mu$m. (B) A confocal scan of a parabolic antenna with a single quantum dot in the center. The scale bar is 1~$\mu$m. (C) Deviation from the parabolic fit of antenna AFM scan.   The scale bar is 1~$\mu$m. }
\label{sfgr:characterization}
\end{figure}

\paragraph{Purcell effect.} We simulated the Purcell effect for a dipole parallel to the glass substrate and at a 10~nm distance from glass interface and in the focal point of parabolic antenna~(Fig.S2). The simulations show no significant change in Purcell factor as the parabolic antenna does not support optical resonances. The curves show in Fig.S2 are for a dipole in the air-glass  interface (blue) and inside the antenna~(red).
\begin{figure}
\includegraphics[width=0.6\textwidth]{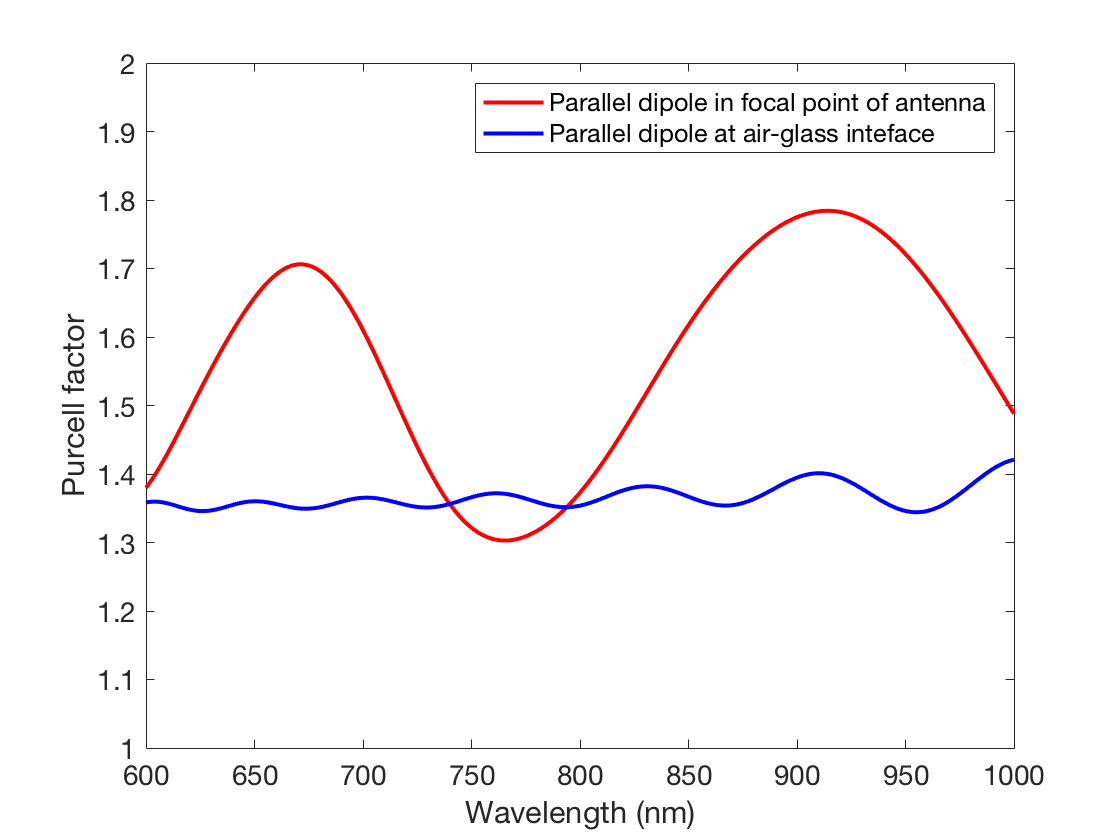}
\caption{\textbf{Purcell effect before and after embedding an emitter in antenna.} A parallel dipole at air-glass interface (blue) and in a parabolic antenna~(red).  }
\label{sfgr:purcell}
\end{figure}

\paragraph{Antenna-emitter alignment.} We align the focal plane of parabolic antenna with a quantum dot by controlling exposure dose and defocussing distance. 
During the alignment, we measured the angular radiation patterns of a quantum dot positioned below and above the focal plane as shown in Fig.S3, which present patterns with varying degree of collimation. 
We characterize the geometrical shape of fabricated antennas by radius to height ratio (RH), which for the case of an emitter in focal point should be equal to 2. 
The radius and height are obtained from AFM scans of the fabricated antennas.
Only when the quantum dot is in the focal plane, i.e. for R:H=2 its emission is fully collimated.
\begin{figure}[H]
\includegraphics[width=0.96\textwidth]{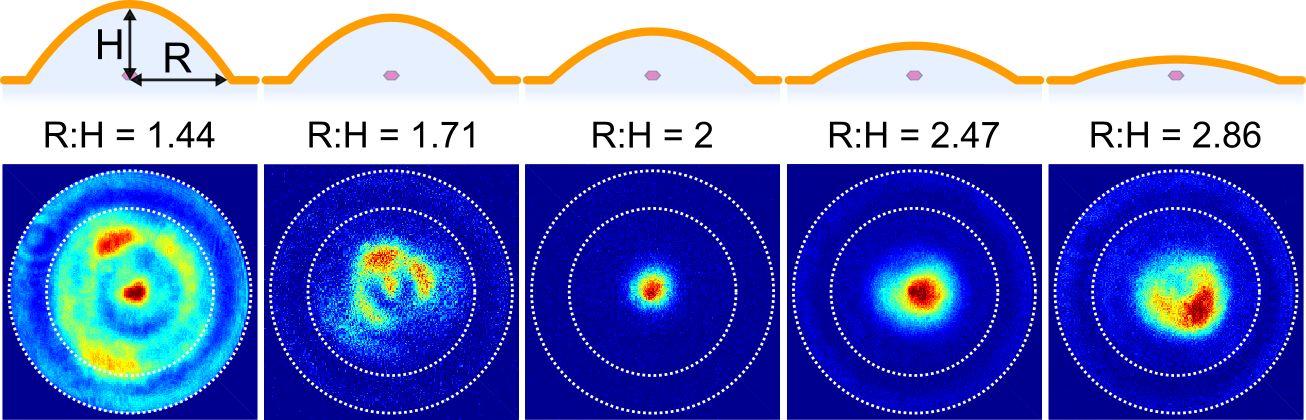}
\caption{\textbf{A quantum dot is out of focal plane of a parabolic antenna.} Experimental angular radiation patterns of quantum dots placed below (R:H\textless2) and above (R:H\textgreater2) the focal plane of a parabolic antenna. The inner and outer dashed white circles represent the air-glass critical angle and the largest collection angle with NA=1.45 objective, respectively. }
\label{sfgr:axial}
\end{figure}

\newpage
\paragraph{Comparison of simulated and experimental radiation patterns.}
The simulated and measured angular radial profiles (extracted from Fig.3) as a function of NA of collection optics are shown in Fig.S4 for an emitter at air-glass interface and in the parabolic antenna.
The experimental results are shown in red, while green and blue curves represent the simulations of the parallel and perpendicular dipole orientations (see legend).

\begin{figure}
\includegraphics[width=0.9\textwidth]{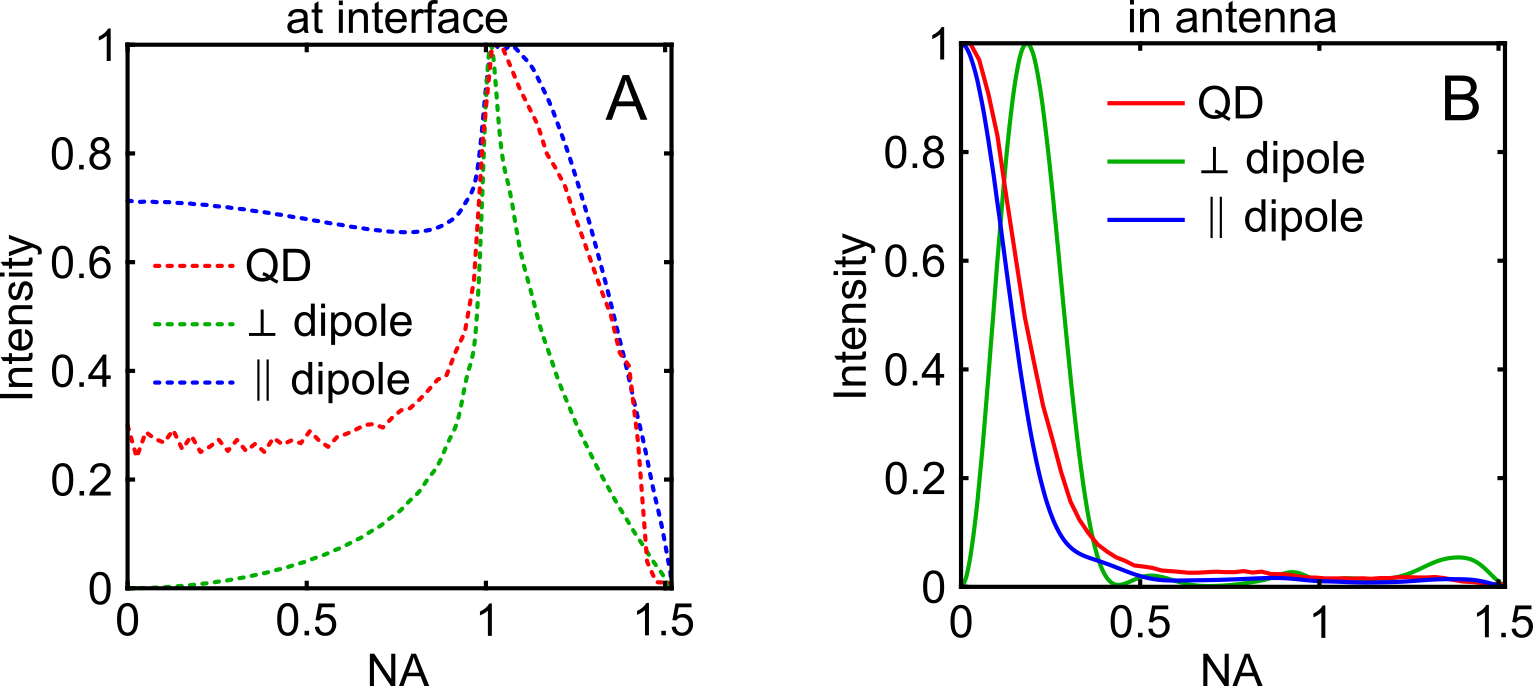}
\caption{\textbf{Radial profiles of measured and simulated angular radiation patterns.} Radially averaged radiation patterns of a single quantum dot at air-glass interface (A) and in the parabolic antenna (B) as a function of numerical aperture NA.}
\label{sfgr:aver}
\end{figure}

\newpage
{
\paragraph{Broadband operation of a metal-dielectric parabolic antenna.}
We simulated the dependencies of collection efficiency and directivity on size of antenna (Fig.S5). Fig.S5 demonstrates the broadband operation of the antenna over the visible part of spectrum.  We have chosen $H=\lambda$ (which corresponds to 3.2~$\mu$m antenna dish opening) as it is the smallest antenna which can direct more than 45\% of the emission into a single-mode fibre (NA=0.2) without any additional collection optics.  
}
\begin{figure}
\includegraphics[width=0.8\textwidth]{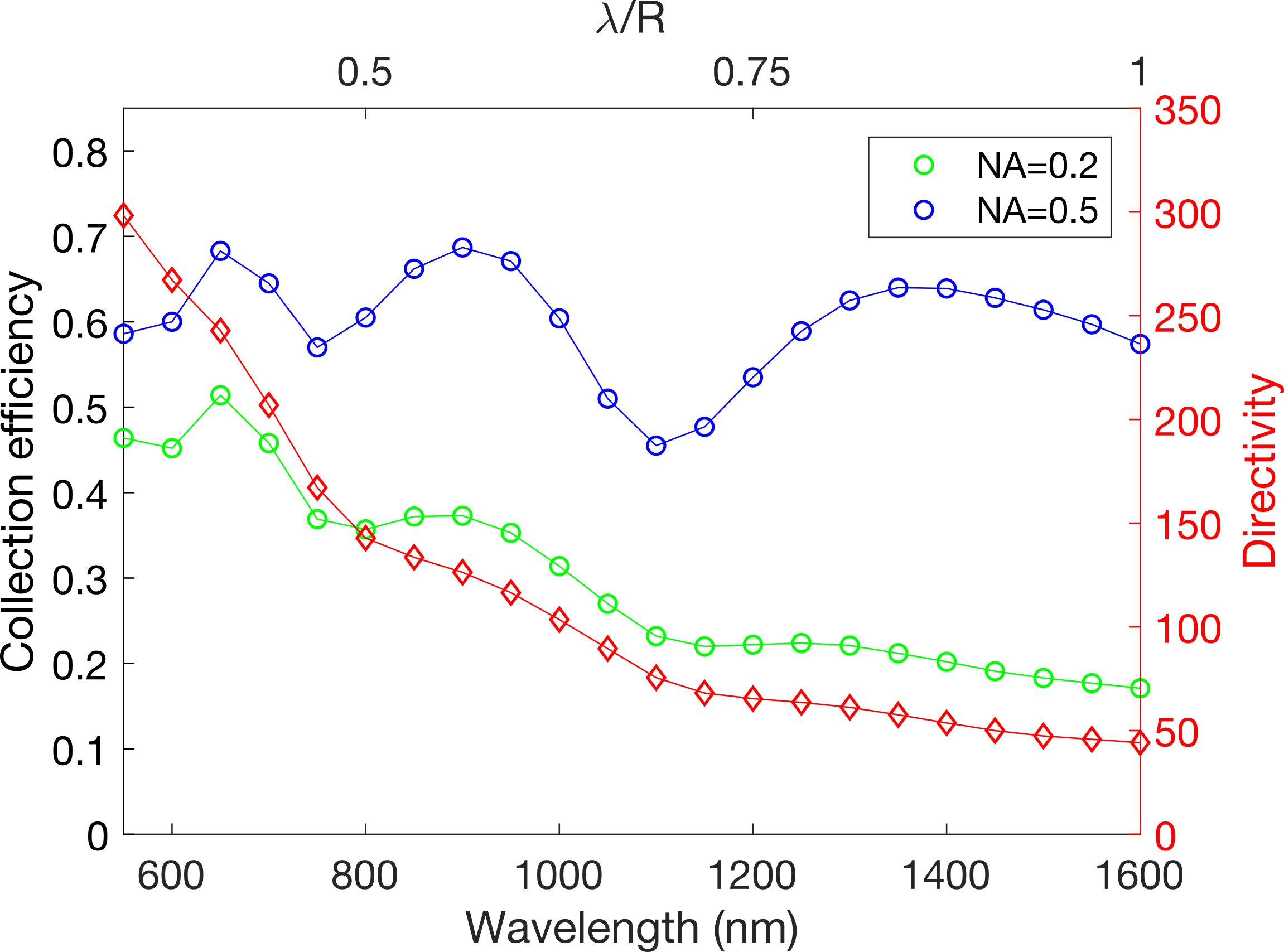}
\caption{{\textbf{Broadband operation of a metal-dielectric parabolic antenna.} Simulated collection efficiency with NA=0.2 (green) and NA=0.5 (blue)  optics, and directivity (red) dependencies on wavelength and effective size of antenna.}}
\label{sfgr:broad}
\end{figure}